\documentclass[conference]{IEEEtran}
\IEEEoverridecommandlockouts
\usepackage[T1]{fontenc}
\usepackage{times}
\usepackage{mathptmx}

\usepackage{cite}
\usepackage{amsmath,amssymb,amsfonts}
\usepackage{algorithmic}
\usepackage{graphicx}
\usepackage{textcomp}
\usepackage{xcolor}
\usepackage{svg}
\usepackage{booktabs} 
\usepackage{float}
\usepackage{url}
\usepackage[bottom]{footmisc}  
\usepackage[most]{tcolorbox} 
\usepackage{enumitem} 
\usepackage{threeparttable}
\usepackage{fancyvrb}
\usepackage{subfig}
\usepackage{orcidlink}
\usepackage[normalem]{ulem}
 
\def\BibTeX{{\rm B\kern-.05em{\sc i\kern-.025em b}\kern-.08em
    T\kern-.1667em\lower.7ex\hbox{E}\kern-.125emX}}

\begin{document}


\title{Supporting Software Formal Verification with Large Language Models: An Experimental Study}

\author{
\IEEEauthorblockN{Weiqi Wang\orcidlink{0009-0005-8247-5994}, 
Marie Farrell\orcidlink{0000-0001-7708-3877}, 
Lucas C. Cordeiro\orcidlink{0000-0002-6235-4272}, 
Liping Zhao\orcidlink{0000-0001-8556-8655}
}
\IEEEauthorblockA{
University of Manchester, Manchester, UK \\ 
Weiqi.Wang-2@postgrad.manchester.ac.uk, \{ Marie.Farrell, Lucas.Cordeiro, Liping.Zhao \}@manchester.ac.uk \\
}
}

\maketitle

\begin{abstract}
Formal methods have been employed for requirements verification for a long time. However, it is difficult to automatically derive properties from natural language requirements. \textit{SpecVerify} addresses this challenge by integrating large language models (LLMs) with formal verification tools, providing a more flexible mechanism for expressing requirements. This framework combines Claude 3.5 Sonnet with the ESBMC verifier to form an automated workflow.
Evaluated on nine cyber-physical systems from Lockheed Martin, \textit{SpecVerify} achieves $46.5\%$ verification accuracy, comparable to NASA's CoCoSim, but with lower false positives. Our framework formulates assertions that extend beyond the expressive power of LTL and identifies falsifiable cases that are missed by more traditional methods. Counterexample analysis reveals CoCoSim’s limitations stemming from model connection errors and numerical approximation issues.
While \textit{SpecVerify} advances verification automation, our comparative study of Claude, ChatGPT, and Llama shows that high-quality requirements documentation and human monitoring remain critical, as models occasionally misinterpret specifications. Our results demonstrate that LLMs can significantly reduce the barriers to formal verification, while highlighting the continued importance of human-machine collaboration in achieving optimal results.
\end{abstract}

\begin{IEEEkeywords}
Large Language Models, 
Formal Verification, 
Requirements Engineering, 
Bounded Model Checking,
Software Verification,
Safety-Critical Systems
\end{IEEEkeywords}

\section{Introduction}

In recent years, formal verification tools have made significant progress in automation. For example, CBMC (C Bounded Model Checker) \cite{kroening2014cbmc} and ESBMC (Efficient SMT-Based Bounded Model Checker) \cite{gadelha2018esbmc} can convert code into verification models, significantly reducing the tedious manual modeling process that traditional tools, such as NuSMV (New Symbolic Model Verifier) \cite{cimatti1999nusmv} and SPIN (Simple Promela Interpreter) \cite{holzmann1997model}, rely on. CBMC and ESBMC use bounded model checking, which systematically explores implementation-level system behaviors within finite execution steps to verify safety properties, unlike traditional model checkers that require manual construction of abstract system models. Although these tools can automatically verify fundamental safety properties, such as overflow and data race conditions, when encountering specific requirements, manual work is still required to accurately specify suitable verification properties from the natural language requirements. This process prevents domain experts from efficiently exploiting the benefits of formal methods \cite{woodcock2009formal}.

Large Language Models (LLMs) provide an opportunity to automate tasks that are currently manual. This is because LLMs have demonstrated excellent capabilities in generating complex program logic and code, such as in the programming capabilities of GPT 4~\cite{achiam2023gpt}, generating temporal logic specifications~\cite{manas2024tr2mtl,cosler2023nl2spec}, model checking property generation~\cite{fang2024assertllm}, and automatic theorem proving~\cite{xin2024deepseek}. 
These capabilities could help domain experts to  automatically transform natural language requirements into verifiable properties. For example, experts in the aerospace field may lack the knowledge of temporal logic required to successfully use formal verification tools or may be unable to accurately transform requirements into corresponding verification properties. 

Traditional approaches to requirements formalization and verification, such as the FRET-CoCoSim toolchain developed by NASA for aerospace software verification \cite{bourbouh2020cocosim}, follow a methodology based on structured language. The FRET (Formal Requirements Elicitation Tool) \cite{giannakopoulou2020formal} leverages structured natural language to formalize the requirements. FRET then translates these formalized requirements into verification properties in CoCoSim for Simulink models, creating a multi-stage pipeline from natural language to formalized verification conditions. While this approach has shown promise in aerospace applications, it still requires significant manual intervention during the translation stage.

This paper proposes \textit{SpecVerify}\footnote{Our code and benchmarks are publicly available \url{https://github.com/LLM-ReqVerif/LLM-ReqVerif}}, a LLM-aided approach to formalize natural language requirements automatically. We aim to leverage the capabilities of large models to align requirements and code, enabling the verification process, which previously required multiple manual interventions, to be highly automated. By comparing the results with those of past manual verifications, we can determine whether this framework has improved the accuracy and automation of the specification and verification task.

We use the Lockheed Martin Cyber-Physical Systems (LMCPS) benchmark~\cite{mavridou2020ten} to evaluate and validate our proposed approach. 
The benchmark comprises several tasks, including signal processing, finite-state control systems, navigation, and system integration. These systems represent realistic cyber-physical applications where requirements must be precisely specified and verified to ensure safety and reliability in operational environments.

The main contributions of our approach are:

\begin{itemize}
    \item \textbf{Improved automation process:} The \textit{SpecVerify} architecture makes the entire process of requirements analysis and assertion generation highly automated. Relying on  ESBMC, it has higher reliability and fewer false reports compared to CoCoSim when examined on the LMCPS problem set. 
        
    \item  \textbf{Enhancing formal specification capability:} Our approach successfully formulates assertions beyond traditional LTL-based methods. This expansion enables the process to automatically generate auxiliary variables to record the previous state, event counter, and other formulas. 
    \item  \textbf{It contributes resources for cyber-physical systems verification:}  We release our framework, benchmark, and evaluation results publicly to address the lack of C datasets in the research on cyber-physical systems.
\end{itemize}

This research systematically compares LLM capabilities in formal verification tasks with traditional verification tools. Our analysis focuses on state-of-the-art models (Claude 3.5 Sonnet~\cite{anthropic2024claude} and ChatGPT 4o1~\cite{openai_help}).

The remainder of this paper is organized as follows: In Section \ref{sec:Existing Transformation and Verification Approach}, we review existing verification methods, especially NASA's FRET-CoCoSim workflow. Section \ref{sec:framework} introduces our proposed research framework, which primarily describes how LLMs with ESBMC are used for verification tasks and explores the core features. Section \ref{sec:methodology} discusses the research methods, datasets and the evaluation design from verification methods from two dimensions: verification logic and results. Section \ref{sec:results} discusses the effectiveness of \textit{SpecVerify} in completing the task based on the research questions and compares the advantages and disadvantages with traditional workflows. Section \ref{sec:threats} analyzes the validity threats, primarily discussing the research's scalability scope, the limitations of the current study, and the methods to address these limitations in the future. Section \ref{sec:conclusion} summarizes the framework's performance and presents key findings, as well as plans for future work.
\section{Existing Transformation and Verification Approaches}
\label{sec:Existing Transformation and Verification Approach}

\subsection{Current NASA Requirements Driven Verification Workflow}

{This section analyzes existing automated verification approaches, with particular focus on NASA's FRET-CoCoSim pipeline—currently the most advanced toolchain for translating natural language requirements into formal verification models in aerospace verification. We examine both the capabilities and limitations of this workflow to establish our research baseline and identify key automation gaps that SpecVerify aims to address.}

NASA Ames Research Center developed a series of tools to automate the functional verification of aerospace software~\cite{giannakopoulou2020formal,perez2022automated}. They built a workflow that integrates FRET and CoCoSim. First, FRET transforms requirements written in structured natural language into LTL \cite{vardi2005automata} expressions and a simplified version of Lustre code. Then, CoCoSim converts the Lustre code into verification conditions that can be directly attached to Simulink models. These models are verified using a customized version of Kind2~\cite{champion2016kind}, which checks if the model's inference functions meet design requirements. This approach partially automates the complete process from requirements capture to model verification.

\subsubsection{FRET (Capabilities and Automation Gaps):}

{FRET excels at structured requirement formalization with over 500 template keys~\cite{mavridou2020ten} and has been successfully applied in aerospace, medical device~\cite{farrell2024fretting}, and industrial automation~\cite{fink2024verifying} domains.} However, three key barriers limit end-to-end automation~\cite{bourbouh2020integration,giannakopoulou2020formal}:

\begin{enumerate}[label=(\alph*)]
    \item {\textbf{Manual Language Translation:}} The natural language requirements written by engineers must be converted into FRETish (a structured natural language used by FRET). While generally user-friendly, this still poses a barrier to use and may introduce human errors in large-scale system verification as documented in aerospace software studies~\cite{mavridou2020ten,bourbouh2020integration}.
    
    \item \textbf{ Manual Variable Mapping:}  {Abstract requirement variables must be manually mapped to concrete system variables. This can be especially challenging when bridging high-level concepts to implementation details.}
    
    \item \textbf{FRETish Temporal Expression Limitations:} Standard temporal logic cannot express certain verification scenarios, such as: \textit{complex historical dependencies}, \textit{structural constraints}, \textit{optimization requirements}, \textit{event counting over time}.

\end{enumerate}

\subsubsection{CoCoSim{ (Additional Manual Steps})}

Once requirements are formalized in FRET, CoCoSim~\cite{bourbouh2020cocosim} addresses the next stage of the verification pipeline by converting specifications into verification conditions for Simulink models. However, this extension introduces additional manual intervention requirements: secondary variable mapping, manual model construction for complex requirements, and the need for expertise across multiple specialized tools (FRET → Lustre → CoCoSim → Simulink → Kind2).

{ Previous use cases encountered FRET's language expressiveness limitations and they employed auxiliary variables to handle verification scenarios that FRETISH cannot directly express \cite{mavridou2020ten,bourbouh2020integration}. For example, for requirements needing first-order logic or previous value references, they used placeholder auxiliary variables in FRET that were replaced with actual quantifiers during Lustre code generation.}

Regarding the assessment of current state, while the FRET-CoCoSim pipeline represents significant progress in automated verification, our analysis reveals that critical manual intervention points persist at every major stage: natural language translation, variable mapping, and complex requirement handling. These automation gaps, combined with the steep learning curve across multiple specialized tools, limit the approach's scalability for large-scale industrial verification. The workaround strategies employed by previous case studies—while technically sound—further highlight the fundamental limitations in expressiveness and automation that motivated our development of \textit{SpecVerify}.

\subsection{LLM-aided Verification Methods}

Early research focuses on evaluating specification generation quality through expert assessment. nl2spec~\cite{cosler2023nl2spec} demonstrates LLM proficiency in transforming natural language into temporal logic formulas, while PropertyGPT~\cite{liu2024propertygpt} focuses on smart contract verification and AutoSpec~\cite{wen2024enchanting} extends to general-purpose programs. These approaches typically establish ground truth specifications as baselines and measure LLM performance against predefined correctness criteria.

More recent work evaluates LLM-generated specifications through direct tool integration and verification outcomes. SpecGen~\cite{ma2024specgen} generates preconditions, postconditions, and loop invariants, then validates them through Java Modeling Language verification via theorem provers. nl2postcondition~\cite{endres2024can} uses fault injection to improve property generation, successfully detecting 64 historical bugs from the Defects4J benchmark, though its reliance on existing test cases limits discovery to known failure patterns. AssertLLM~\cite{fang2024assertllm} demonstrates hardware assertion generation for Verilog verification tools, requiring LLMs to analyze both design specifications and implementation code. 

The hardware verification domain has shown particularly promising results, with LLM-based approaches successfully generating verifiable assertions~\cite{ma2024verilogreader,kang2025fveval} and properties that integrate well with existing tools, leading to significant excitement in the LLM aided verification community about automated specification generation. Hardware verification benefits from more standardized documentation and well-defined interfaces compared to software systems. These approaches remain confined to specific domains, with most requiring manual validation steps that create scalability bottlenecks. Software verification faces more complex variable mapping challenges, particularly in safety-critical domains where specification errors can have severe consequences.

While current LLM verification research has established basic feasibility, comprehensive end-to-end automation approaches that integrate natural language processing with industrial-strength verification tools remain largely unexplored, particularly for safety-critical software verification workflows.
\section{Proposed Verification Approach}
\label{sec:framework}

\begin{figure*}[htbp]
\centerline{\includegraphics[width=0.8\textwidth, clip, trim=0cm 0cm 0cm 0cm]{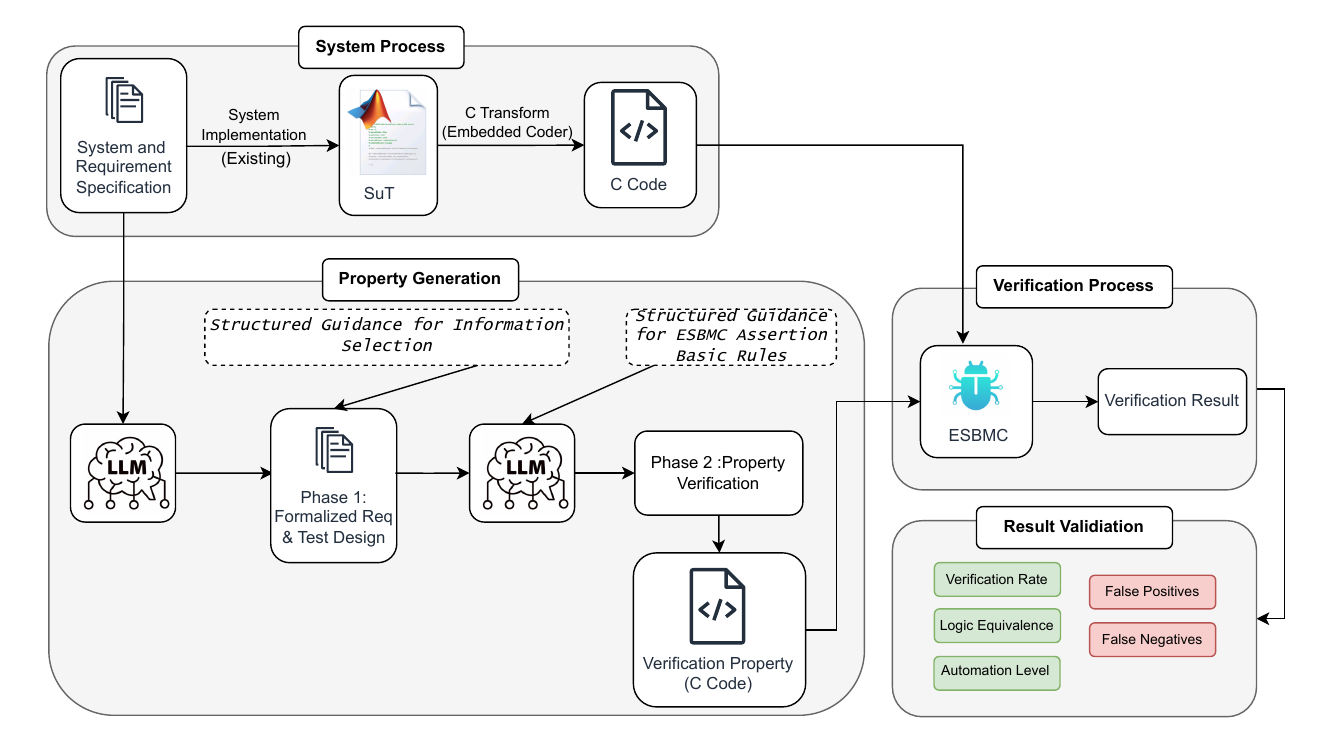}}
\caption{Proposed Approach: Software Verification Process Supported by LLMs. LLMs guide formal specification and property generation, with ESBMC providing automated verification results.
}
\label{fig:1}
\end{figure*}

This section introduces the \textit{SpecVerify} approach, along with its core components. It includes how automatic conversion is conducted and its advantages.


The \textit{SpecVerify} framework utilizes LLMs to establish a direct connection between natural language requirements and C language verification code. As illustrated in Figure~\ref{fig:1}, our framework enhances the traditional formal verification workflow with a two-stage approach:

\textbf{Requirement formalization phase:} A LLM converts natural language requirements into a formalized requirement specification, which replaces the function of FRET. This process involves capturing requirements and mapping variables, including the mapping of high-level requirements to subsystem-level behaviors, variable relationships, and timing logic. The generated formalized requirement specification serves as an intermediate document, which humans can view and confirm to be consistent with the original requirements.

\textbf{Verification property generation phase:} In the second step, the LLM generates C code assertions applicable to the ESBMC specification based on the formal specification. The entire process replaces the role of CoCoSim. These assertions contain all necessary verification conditions and boundary checks.



The \textit{SpecVerify} approach supports the transformation from natural language requirements to $C$ verification code through the following three improvements:

\begin{itemize}
    \item \textbf{Enhanced expressiveness:} Our LLM-driven approach addresses historical dependencies, interactions between multiple variables, non-standardized timing conditional expressions, and non-functional requirements. The subsequent LLM-based code generation tool can generate auxiliary variables at the corresponding location of the code to provide the correct expression.
    \item \textbf{Optimized intermediate representation:} Our framework replaces the conversion to an intermediate representation layer (such as the Lustre or NuSMV models) with a formal specification. The document retains the meaning of the requirements and attempts to present the verification plan in a user-readable format with actual mapping.
    \item \textbf{LLM-driven variable mapping:} Our approach utilizes the semantic understanding capabilities of LLMs to convert abstract requirements into detailed relationships and lower-level variables directly. This approach eliminates the manual variable mapping step in traditional processes, thereby increasing the automation of the validation process.
\end{itemize}

\section{Experimental Design}
\label{sec:methodology}

\subsection{Research Questions}

Our experimental design is guided by the following key research questions (RQs):
\begin{tcolorbox}[colback=blue!5, colframe=blue!75, title=Research Questions, fonttitle=\bfseries, before skip=10pt, after skip=10pt]

\begin{itemize}
\small
\item[\textbf{RQ1:}] How does \textit{SpecVerify} automatically transform requirements into properties for formal verification?
\begin{itemize}
\item What does the current workflow look like?
\item Is it functionally equivalent to traditional tools?
\end{itemize}
\item[\textbf{RQ2:}] How does \textit{SpecVerify} perform compared to traditional tools?
\begin{itemize}
    \item What causes the differences?
    \item How can we evolve to address current challenges?
\end{itemize}
\end{itemize}

\end{tcolorbox}
In the following sections, we describe the core components of our experimental study.

\subsection{Dataset}
\label{Overview of Research Content and Dataset}
\begin{table*}[htbp]
\caption{Lockheed Martin Cyber-Physical System Tasks Overview \cite{mavridou2020ten}}
\centerline{\includegraphics[width=\textwidth, clip, trim=1cm 2cm 1cm 1cm]{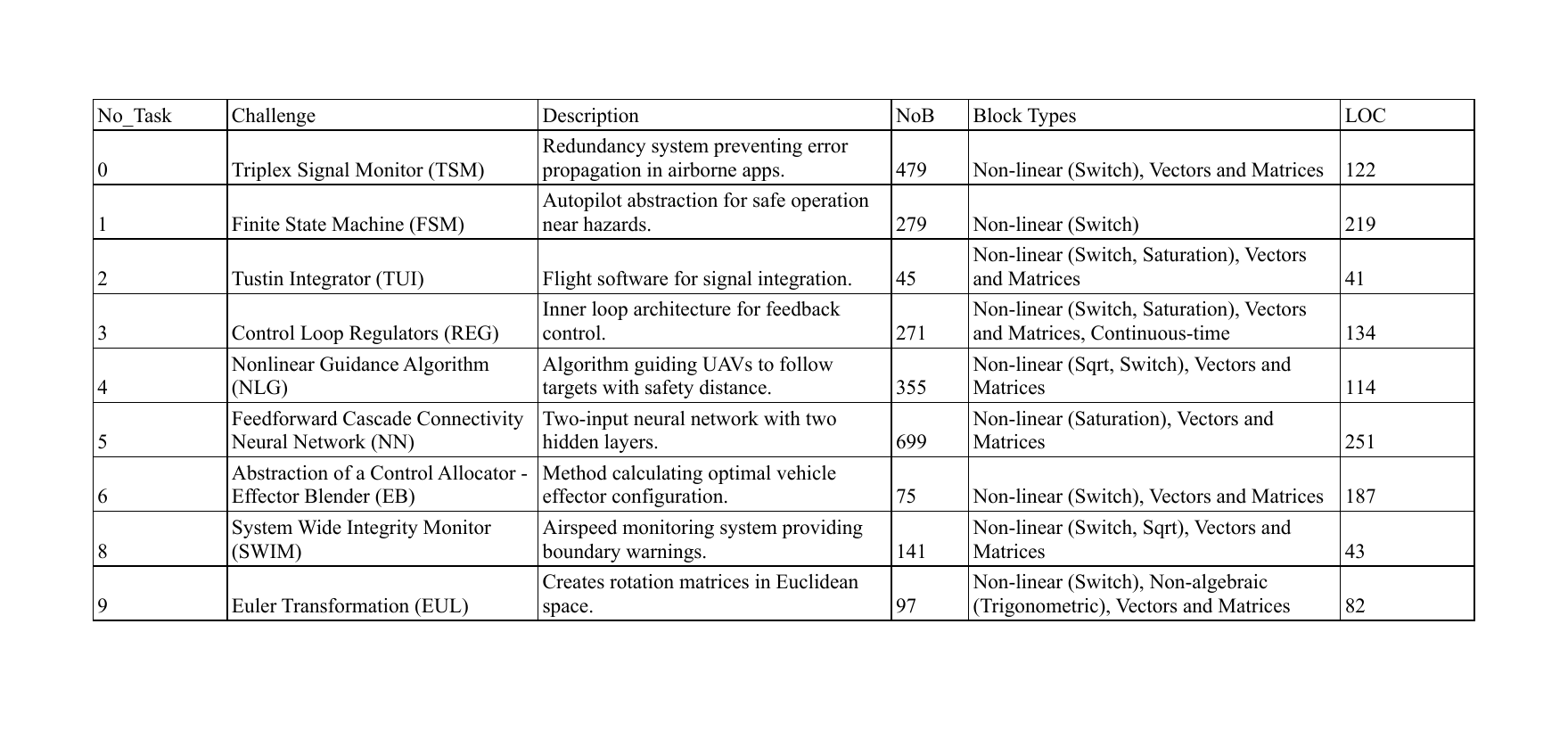}}
\begin{tablenotes}
\small
\item NoB: Number of Blocks, indicating system complexity.
\item LOC: Lines of Code, representing implementation size.
\item This table summarizes nine aerospace control system components used in flight control and monitoring applications.
\end{tablenotes}
\label{Table of the LMCPS}
\end{table*}

The LMCPS benchmark consists of a series of Simulink-related aerospace examples developed by Lockheed Martin. There is limited open-source software. However, this benchmark addresses many issues in embedded system development, which we will demonstrate below. We use Embedded Coder to convert Simulink models into C code which is the input for ESBMC.

\textit{Benchmark Composition:}
As shown in Table~\ref{Table of the LMCPS}, the benchmark comprises nine distinct challenges across four primary domains (one of the ten original datasets failed to be converted to C code due to an out-of-date component in Simulink):

\begin{itemize}
\item Signal Processing and Monitoring (TSM, TUI): Real-time signal analysis and anomaly detection in sensor data streams.

\item Finite State Control Systems (FSM, REG): Supervisory control of states to achieve system behavior management.

\item Navigation and Guidance (NLG, NN, EB): A series of complex trajectory calculation and navigation algorithms.

\item System Integration and Safety (SWIM, EUL): Encompasses system-level safety properties and integration verification, ensuring overall system reliability.

\item Task 7, autopilot, was removed since the model failed to be converted to C code, as we could not find the library that was used. 

\end{itemize}





\subsection{Establishing baseline facts}
We collected the CoCoSim results for these models, including those based on customized Kind2 and SLDV, and then compiled our results. If any differences were found, we generated test cases based on the counterexamples. We executed the program with additional print statements to directly observe the runtime behavior, allowing us to verify that the results were consistent and determine whether ESBMC had identified a new error.

\begin{itemize}
\item 12 requirements were successfully verified as provable, demonstrating that they satisfied the specifications.
\item 17 cases were proved falsifiable, demonstrating there is at least one counterexample that does not meet the requirements.
\item 29 cases remained undetermined ($39\%$), mainly because of the complexity of the calculation.
\end{itemize}

This distribution represents the challenge for formal method tools in the real world, where some undetermined cases still exist for formal methods to explore, making this benchmark realistic and valuable for evaluating novel verification approaches.

\subsection{Experimental Setup}

We conducted the experiments on a Dell laptop equipped with an Intel Core i7-9750H CPU, 16 GB of DDR4 RAM, and a 1 TB SSD, running Ubuntu 24.04.1 LTS. We leveraged two online large-scale language models: Claude 3.5 Sonnet (Anthropic) and ChatGPT 4.0 (OpenAI). We used ESBMC v7.7 as the verification engine.

We initially selected three LLMs as the models for our research. We selected ChatGPT because it was one of the earliest LLMs to attempt code analysis and generation~\cite{chen2021evaluating} and ranked among the top in multiple code benchmarks~\cite{jimenez2023swe,aiderLeaderboard}. At the same time, Claude was chosen as the current state-of-the-art code generation model according to the Aider LLM benchmark \cite{aiderLeaderboard}, so we selected it as another comparison model. We initially included Llama 3.1 8B~\cite{meta2024llama} as a local model for comparison. Still, its results were excluded from our study as it failed to perform any tasks in our research due to the significant size difference compared to the larger models. 

For each LMCPS benchmark requirement, we followed a consistent process. We provided the natural language requirement and corresponding C code to the LLM. The LLM then generated a semi-formal specification and translated it into C assertions. We integrated these assertions into the code and verified them using ESBMC. Finally, we compared the results with the ground truth using our dual validation approach.

\subsection{Evaluation of Results}
We constructed the verification process through two methods to ensure the reliability and correctness of LLM-driven verification:

\subsubsection{Automation level evaluation (for RQ1)}
To answer RQ1’s question about the automated conversion process of \textit{SpecVerify}, we compared and analyzed the automation levels of different tools:

\begin{itemize}
\item \textbf{Manual intervention point identification}: CoCoSim's research recorded several manual intervention points that LTL cannot express. We compared it with the SpecVerify workflow to see if this process can be automated.
\item \textbf{Variable mapping analysis}: Evaluate the way and accuracy of each tool to map abstract variables in natural language requirements to code variables.
\item \textbf{Language constraint comparison}: Analyze the information of each tool at the manual point to determine why the new method may have advantages and disadvantages.
\end{itemize}

\subsubsection{Logical comparison (for RQ1)}

We want to verify whether the LLM and manual verification are logically equivalent.

\begin{itemize}
    \item Our implementation utilizes a 3-tuple structure, similar to CoCoSim (precondition--function--postcondition), which facilitates comparison.
    \item This format simplifies comparison, allowing only each set of conditions and results, as well as the definition of additional formulas, to be compared.
    \item Currently, different languages are used, resulting in a lack of automation, and considerable manual work is required to verify the verification code generated by Claude.
    \item The code generated by ChatGPT is excluded from the extensive comparison due to its poor performance and limited practicality in verifying the value in this context.
\end{itemize}

\subsubsection{Result Comparison (for RQ2)}

We established a comparison mechanism between the verification results of ESBMC and those of traditional tools, including CoCoSim and SLDV. The process follows these steps:

\begin{itemize}
    \item First, we collect verification results and generate test case-based witnesses with counterexamples of different results.
    \item These witnesses are written into the main class and put into the original code through execution to determine whether they would cause actual problems.
    \item If the error is not triggered, we return to the counterexample and combine it with the original code to determine where the error comes from. If it still cannot be solved, we combine the solver source code to determine whether there is an error.
    
\end{itemize}

\section{Experimental results and analysis}

\label{sec:results}

This section first presents the experimental results for each research question (RQ1 and RQ2), and then provides an analysis of these results.


\subsection{Results for RQ1}

\begin{table}[ht]

\centering

  \caption{Automation of workflow}
\resizebox{\columnwidth}{!}{%
  \begin{tabular}{lcccc}
  \hline
  \textbf{Aspect} & \textbf{CoCoSim}  & \textbf{SLDV} & \textbf{Claude 3.5} & \textbf{ChatGPT4o1} \\
  \hline
  Automation Level & Semi & Semi & \textbf{Full} & \textbf{Full} \\
  Manual Mapping & Required & Required & \textbf{Not Required} & \textbf{Not Required} \\
  Language Constraints & Fixed & Fixed & \textbf{Flexible} & \textbf{Flexible} \\
  \hline
  \label{tab:comparative_analysis}
  \end{tabular}%
}

\end{table}

\begin{table}[!ht]
\caption{Logic Equivalence Analysis Results}
\label{Logic Equivalence Analysis Results}
\begin{center}
\begin{tabular}{lrr}
\toprule
Category & Count & Percentage \\
\midrule
Logic Equivalent & 46 & $79.31\%$ \\
LLM Misunderstanding & 2 & $3.45\%$ \\
LLM Lacking Assumption & 2 & $3.45\%$ \\
Benchmark Skipped & 4 & $6.90\%$ \\
LLM Sequence Reversal & 2 & $3.45\%$ \\
Over Verification (LLM) & 1 & $1.72\%$ \\
Over Verification (CoCoSim) & 1 & $1.72\% $\\
\bottomrule
\end{tabular}
\end{center}
\end{table}

In Table~\ref{Logic Equivalence Analysis Results}, benchmark skipping refers to the fact that the verification team was unable to construct verification properties due to the limited expressiveness of LTL. Over-verification refers to verifying content that exceeds the original verification requirements. Sequence reversal occurs in the calculation case, and the two cases are the same. The LLM incorrectly generates the opposite order that assertion one responds to requirement two, while assertion two responds to requirement one. 

To better understand the nature of these discrepancies and their implications for automated verification, we examine detailed case studies that illustrate both successful transformations and problematic cases encountered in the LLM-based requirement interpretation process.

\subsubsection{Case Study 1 (Hoare Tuple Logic Analysis)}

To illustrate how LLMs transform natural language requirements into formal verification properties, we examine a representative case where our framework converts implementation-specific requirements into Hoare tuple logic~\cite{pratt1976semantical}. This analysis demonstrates the structural differences between traditional Lustre specifications and LLM-generated formal requirements.

\begin{table}[h]
\centering
\caption{Information Content Comparison between Lustre and LLM Specifications.}
\label{tab:hoare_comparison}
\begin{tabular}{p{1.5cm}p{2.7cm}p{2.7cm}}
\toprule
\textbf{Component} & \textbf{Lustre Specification} & \textbf{LLM Formal Requirement} \\
\midrule
\textbf{Precondition} & \texttt{var no\_fail : bool = (FC =0)} & \texttt{rtDW.Delay1\_DSTATE [2] == 0} \\
& & \texttt{(No-fail state)} \\
\midrule
\textbf{Variables} & \texttt{mid\_value: real = if...} & \texttt{sel\_val = mid-value of (ia, ib, ic)} \\
\midrule
\textbf{Postcondition} & \texttt{set\_val = mid\_value} & \texttt{sel\_val = mid-value} \\
\midrule
\end{tabular}
\end{table}

The comparison in Table~\ref{tab:hoare_comparison} reveals several key insights about LLM-based formal verification. While traditional Lustre specifications focus on abstract mathematical relationships, LLM-generated requirements maintain closer ties to implementation details. The LLM approach includes specific variable names (\texttt{rtDW.Delay1\_DSTATE[2]}) and implementation context, making the verification properties more accessible to domain experts who are familiar with the actual code structure.

This approach demonstrates both the strengths and limitations of LLM-based verification. While it reduces the abstraction barrier between requirements and implementation, it may introduce implementation-specific dependencies that could affect the generalizability of verification properties.

The following case studies examine specific instances where LLMs encountered difficulties in requirement interpretation, focusing on the two cases of misunderstanding and the two cases of lacking assumptions identified in our logic equivalence analysis.

\subsubsection{Case study 2 (Triple Redundancy Voting Miscompare)}

The LLM did not misunderstand the meaning of the document. However, the document's quality was problematic, and its description of triple-redundant voting was oversimplified and partially incorrect. Our current framework does not recognize this type of mismatch and requires clarification.

Here is a description of the miscompare in the document:
\begin{quote}
\textit{Errors will appear as differences in the values of the input set signals. This difference is called a miscompare.}
\end{quote}

However, the requirements document and code implement a triple-redundant voting system. When one signal differs from the other two, the system uses the majority value as the correct output and marks the source of the differing signal as a fault.

\subsubsection{Case Study 3 (System Wide Integrity Monitor)}

The CoCoSim team manually added an assumption that does not exist in the document: that the weight of the aircraft is larger than zero. Whether an assumption should be added is suspicious because it is not stated in advance in the document. However, this also highlights the difference between humans and LLMs: LLMs are unable to improve, expand, or modify the original requirement in the document.

\begin{tcolorbox}[colback=blue!5, colframe=blue!75, title=Findings for RQ1, fonttitle=\bfseries, before skip=10pt, after skip=10pt]
\footnotesize
\textbf{RQ1:} How does \textit{SpecVerify} automatically transform requirements into properties for formal verification? 

\vspace{0.2cm}

\textbf{Answer: }\textit{SpecVerify} achieves full automation in transforming natural language requirements into formal verification properties, leveraging large language models to eliminate manual variable mapping and language constraints required by traditional tools. 

\vspace{0.1cm}

With $79.31$\% logical equivalence to manually created properties, the approach significantly reduces the expertise barrier for implementing formal verification in safety-critical systems, despite occasional misunderstandings and sequence errors. These results answer \textbf{RQ1} by demonstrating successful automation of the requirements transformation process.
\end{tcolorbox}

\subsection{Results for RQ2:}
\begin{figure}[htbp]

    \centering
    \subfloat[Venn Diagram for Claude+ESBMC]{
        \includegraphics[width=0.35\textwidth]{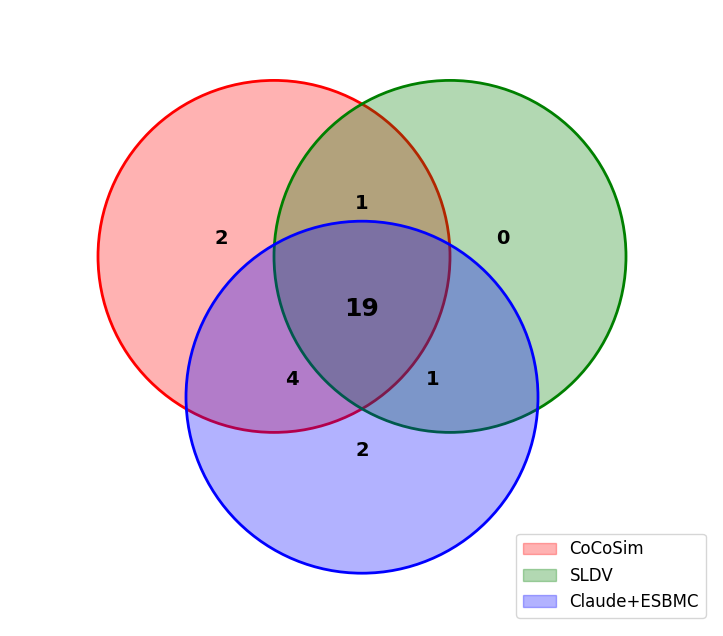}
        \label{fig:venn1}
    }

    \subfloat[Venn Diagram for ChatGPT+ESBMC]{
        \includegraphics[width=0.355\textwidth]{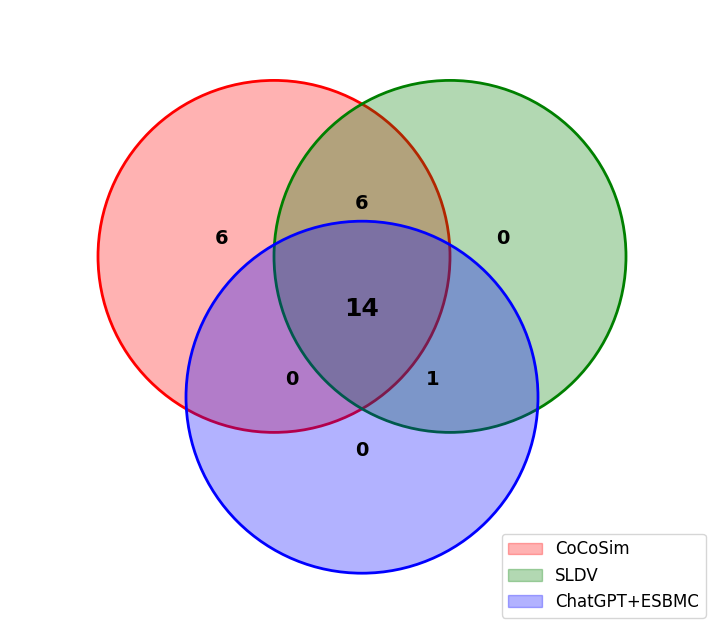}
        \label{fig:venn2}
    }
    \caption{Venn Diagrams comparing verification results}
    \label{fig: Venn}
\end{figure}

We compare our tool with traditional formal verification tools in terms of performance and accuracy. First, we show the verification results of the tools for different task categories. Then, we analyze specific cases to determine whether LLMs or verifiers make the difference.

\begin{table}[htbp]
\centering
\caption{Performance Analysis - Verification Effectiveness by Task Category.}
\label{tab:PerformanceAnalysisNew}
\footnotesize
\begin{tabular}{lcccc}
\hline
\textbf{Task Category} & \textbf{CoCoSim} & \textbf{SLDV} & \textbf{Claude} & \textbf{ChatGPT} \\
 & & & \textbf{+ESBMC} & \textbf{+ESBMC} \\
\hline
\textbf{\textit{Signal Processing}} & & & & \\
TSM & 3/4/4 & \textbf{4/4/4} & 3/4/4 & 2/4/4 \\
TUI & 3/3/5 & 3/3/5 & \textbf{4/5/5} & 0/0/5 \\
\textbf{\textit{Finite State Control}} & & & & \\
FSM$^\dagger$ & \textbf{13/13/13} & \textbf{13/13/13} & \textbf{13/13/13} & \textbf{13/13/13} \\
REG & \textbf{5/10/10} & 0/10/10 & \textbf{5/10/10} & 0/0/10 \\
\textbf{\textit{Navigation}} & & & & \\
NLG & 0/7/7 & 0/7/7 & \textbf{1/7/7} & 0/0/7 \\
NN & 0/4/4 & 0/4/4 & 0/4/4 & 0/4/4 \\
EB & 0/3/5 & 0/3/5 & 0/5/5 & 0/5/5 \\
\textbf{\textit{System Integration}} & & & & \\
SWIM & \textbf{2/2/2} & 1/2/2 & 1/2/2 & 0/1/2 \\
EUL & \textbf{1/8/8} & 0/8/8 & 0/8/8 & 0/8/8 \\
\hline
\textbf{Performance Metrics} & & & & \\
Verified/Formed/Total & \textbf{27/54/58} & 21/54/58 & \textbf{27/58/58} & 15/35/58 \\
Verification Rate (\%) & \textbf{46.5} & 36.2 & \textbf{46.5} & 25.9 \\
False Positives$^\star$ & 2 & \textbf{0} & \textbf{0} & 8 \\
False Negatives$^\star$ & 6 & \textbf{0} & 2 & 2 \\
\hline
Assertion Errors & \textbf{0} & \textbf{0} & \textbf{0} & 23 \\
sin/cos approx. error & 6 & \textbf{0} & \textbf{0} & \textbf{0} \\
\hline
\multicolumn{5}{l}{\scriptsize $^\dagger$FSM: Training cases used for all tools.} \\
\multicolumn{5}{l}{\scriptsize $^\star$ CoCoSim: FN include FN-Unknown due to sin/cos solver error;} \\
\multicolumn{5}{l}{\scriptsize $^\star$ ChatGPT+ESBMC: FP includes FN-Unknown due to non-det input setup assertion.} \\
\end{tabular}
\end{table}

As shown in Table~\ref{tab:PerformanceAnalysisNew}, our experiments evaluated the performance of the CoCoSim, SLDV, and LLM-based verification methods across various task types, including methods and traditional techniques. Fig~\ref{fig: Venn} shows that Claude+ESBMC finds two extra errors than CoCoSim and SLDV could discover, while ChatGPT+ESBMC finds a non-competitive number of errors.

Regarding the precise floating-point operations of ESBMC, we identified floating-point errors that CoCoSim did not detect, which were confirmed by writing and executing test cases. 
\begin{align}
a &= 1.813356e+24f \label{eq:value_a}\\
  &= \texttt{01100111 10111111 11111111 00011010}_2\nonumber\\[0.1em]
b &= 2.328307e-10f \label{eq:value_b}\\
  &= \texttt{00101111 10000000 00000000 00000010}_2\nonumber\\[0.1em]
c &= 1.999512e+0f \label{eq:value_c}\\
  &= \texttt{00111111 11111111 11110000 00000000}_2\nonumber
\end{align}
The correct median is $c = 1.999512$ from equation~\eqref{eq:value_c}. The algorithm computes the mean as shown in equation~\eqref{eq:mean_formula}:
\begin{equation}
\mu = \frac{a + b + c}{3} \label{eq:mean_formula}
\end{equation}
In IEEE 754 floating-point arithmetic:
\begin{itemize}
\item Values $b$ and $c$ from equations~\eqref{eq:value_b} and~\eqref{eq:value_c} \textit{are effectively lost during addition with $a$ (equation~\eqref{eq:value_a}) due to their relatively tiny magnitude}
\item $a + b + c \approx a$
\item $\mu \approx \frac{a}{3} \approx 6.044520e+23$
\end{itemize}
When computing and comparing the distances as defined in equations~\eqref{eq:dist_a}, \eqref{eq:dist_b}, and~\eqref{eq:dist_c}:
\begin{align}
|a - \mu| &\approx 1.208904e+24 \label{eq:dist_a}\\
|b - \mu| &\approx 6.044520e+23 \text{ (since $b \approx 0$ relative to $\frac{a}{3}$)} \label{eq:dist_b}\\
|c - \mu| &\approx 6.044520e+23 \text{ (since $c \approx 0$ relative to $\frac{a}{3}$)} \label{eq:dist_c}
\end{align}

Due to floating-point rounding errors, the algorithm incorrectly selects $b$ as the median value (the value closest to the mean computed in Eq.~\eqref{eq:mean_formula}) when the actual median should be $c$ based on the proper ordering $b < c < a$. This error remained undetected in CoCoSim's rational number verification. Our ESBMC-based tool, which uses floating-point semantics, successfully identified this error. The correct implementation uses direct comparison, avoiding floating-point arithmetic issues as described in Eq.~\eqref{eq:correct_impl}:
\begin{equation}
\text{mid} = \max(\min(a, b), \min(\max(a, b), c)) \label{eq:correct_impl}
\end{equation}

\subsubsection{Case Study 4 (Avoiding One Human Error)}

During our research, we discovered a human error in the CoCoSim study. In the REG-003 verification component, the team mistakenly connected the logic that should have been linked to the "$+1$" operation to a constant $0$

The error occurs in the following code segment:
\begin{Verbatim}[fontsize=\small,frame=single]
var result = if (input > 50.0) 
    then zero // counter should increase 1
    else zero; // constant 0 
\end{Verbatim}
This error will cause the verification tool to ignore cases where \texttt{input} exceeds 50.0, potentially missing critical safety violations. Our tool avoids similar mistakes, illustrating the benefit of automating variable mapping.

\begin{tcolorbox}[colback=blue!5, colframe=blue!75, title=Findings for RQ2, fonttitle=\bfseries, before skip=10pt, after skip=10pt]
\footnotesize
\textbf{RQ2:} How does \textit{SpecVerify} perform compared to traditional tools? 

\vspace{0.2cm}

\textbf{Answer: }\textit{SpecVerify} achieved comparable verification rates to traditional tools while detecting additional floating-point errors, which answers RQ2: although it introduces new challenges in document interpretation, empirical evidence suggests competitive performance with enhanced error detection capabilities.
\end{tcolorbox}
\section{Study Limitations}
\label{sec:threats}

When evaluating our research on LLM-assisted formal verification, the following external and internal factors have cast a shadow on its effectiveness and scalability. We also propose some roadmaps to address these issues in the future.


\subsubsection{Dataset Coverage}
The main limitation of our experimental data includes the scope of use and sample size. Our evaluation utilized C code generated by Simulink, ranging from $41$ to $251$ lines, which may not fully represent real-world systems, as these often contain thousands of lines of code and exhibit more complex interactions. The current validation tasks with Lockheed Martin mainly include aircraft-related tasks. Additionally, code automatically generated from Simulink models may be more structured than manually written code, which can limit the generalizability of the results. To overcome this limitation, our future work will expand our benchmark to include diverse real-world codebases of varying scales, incorporate manually written code samples, and increase requirement diversity across different verification domains. 

\subsubsection{Model Selection Considerations}
Our tool selection, comparing Llama 3.1 8B (8 billion parameters) with Claude 3.5 and ChatGPT 4 (over 100 billion parameters), creates an uneven baseline. A more appropriate approach would involve comparing models of similar scale, such as Llama 70B, or other recent large-scale open models like DeepSeek R1 \cite{guo2025deepseek}.

\subsubsection{Assessment Framework Limitations}

The validation of our framework’s results faces two key limitations. Our scoring system may not fully reflect the strengths of different tools, such as the precise and efficient computation of ESBMC and the task-specific computational optimization of CoCoSim. In addition, the contributions of LLMs and ESBMC are mixed, making it necessary to manually verify and analyze the results to determine what leads explicitly to improved or reduced performance.

\subsubsection{Logical Representation Challenges}

Using triples (precondition, function, postcondition) for logical equivalence evaluation may oversimplify the complex logical relationships in large-scale verification scenarios. This may result in verification properties that appear roughly consistent but are pretty different. Although there is no ideal solution at present, the involvement of human evaluators in the evaluation process introduces the risk of subjective bias and inconsistent judgment.

\section{Conclusion}
\label{sec:conclusion}
Our experiments revealed notable performance differences between language models in generating verification properties from requirements documents. ChatGPT 4o1 exhibited significant issues with code generation, often producing assertions that failed to compile due to syntax errors or incorrect variable references. Claude 3.5 proved more reliable in generating syntactically correct verification code consistently. We also found that specification quality significantly impacts results—when requirements contained ambiguities or were poorly formalized, both models struggled to generate adequate verification properties. This reinforces the importance of rigorous specification practices in requirements engineering for formal software verification workflows.

Through our evaluation, we observed that while the automation capabilities show promise, some misinterpretations inevitably arise when requirements documents contain ambiguities or underspecified constraints. In traditional verification workflows, such misunderstandings are typically resolved through iterative discussions between verification engineers and domain experts. Our framework generates intermediate formalized specifications and test designs that can surface these interpretation issues early; however, our current fully automated approach processes these documents directly into verification properties without human review, thereby missing opportunities to identify and correct misunderstandings before they propagate through the verification workflow.

The framework achieves a 46.5\% property verification completion rate, with incomplete cases primarily attributed to bounded model checking limitations—specifically state space explosion and depth bounds—rather than property formalization errors. While this performance is comparable to NASA's CoCoSim framework in terms of overall verification completion rate, \textit{SpecVerify} demonstrates superior reliability through ESBMC integration: producing 2 fewer false positives and 6 fewer false negatives compared to CoCoSim, indicating improved accuracy in verification outcomes.

The majority of incomplete verification cases resulted from solver timeouts rather than incorrect property formalization or logical errors in the verification process. For cases where bounded model checking encounters state space explosion or decidability constraints, we are investigating program abstraction techniques. However, such abstractions require additional checks to ensure the simplified model still accurately captures the original system's behaviour.

Future work will expand beyond Simulink-generated code to encompass diverse real-world systems, develop interactive disambiguation mechanisms for resolving specification ambiguities, and aims to integrate dynamic test case generation for complex navigation scenarios. \textit{SpecVerify} represents progress toward democratizing formal verification for domain experts; however, substantial advances in human-AI collaboration remain necessary before achieving truly autonomous verification capabilities in safety-critical systems.










\clearpage
\bibliographystyle{IEEEtran}


\end{document}